\documentclass[aps,twocolumn,superscriptaddress,prl]{revtex4-2}
\usepackage[utf8]{inputenc}

\usepackage{graphicx}
\usepackage{geometry}
\usepackage{amsmath, amssymb} 
\usepackage{breqn} 
\usepackage{empheq,etoolbox} 
\usepackage{nomencl} 
\makenomenclature
\usepackage{xargs}                      
\usepackage[pdftex,dvipsnames]{xcolor}  

\usepackage{hyperref}

\newcommand{\ladhyx}{LadHyX, CNRS, Ecole Polytechnique, Institut Polytechnique de Paris, 91120 Palaiseau, France}
\newcommand{\pasteur}{Institut Pasteur, Université Paris Cité, Physical microfluidics and Bioengineering, 25-28 Rue du Dr Roux, 75015 Paris, France}

\newcommand{\Qbyp}{Q_{\rm byp}}
\newcommand{\Qthru}{Q_{\rm thru}}

\newcommand{\Rbyp}{R_{\rm byp}}
\newcommand{\Rthru}{R_{\rm thru}}
\newcommand{\Rgel}{R_{\rm gel}}

\begin{document}

\title{Nonlinear clogging of a rectangular slit by a spherical soft particle}

\author{Charles Paul Moore}
\affiliation{\pasteur}
\affiliation{\ladhyx}
\author{Julien Husson}
\affiliation{\ladhyx}
\author{Arezki Boudaoud}
\affiliation{\ladhyx}
\author{Gabriel Amselem}
\affiliation{\ladhyx}
\author{Charles N. Baroud}
\affiliation{\pasteur}
\affiliation{\ladhyx}

\date{\today}

\begin{abstract}
The capture of a soft spherical particle by a rectangular slit leads to a non-monotonic pressure-flow rate relation at low Reynolds number. In the presence of the trapped particle the flow-induced deformations focus the streamlines and pressure drop to a small region. This increases the resistance to flow by several orders of magnitude as the driving pressure is increased. As a result two regimes are observed: a flow-dominated regime for small particle deformations, where flow rate increases with 
pressure, and an elastic-dominated regime in which solid deformations block the flow. 
\end{abstract}

\maketitle

Solid deformations can have a major influence on fluid flow in the regime of strong fluid-structure coupling. A wide range of complex behaviors, including oscillations that lead to catastrophic failures or blockages, can be observed when the inertial effects of the fluid and solid couple together~\cite{paidoussis2010fluid}. Recently an increased attention has been paid to the effet of a low Reynolds number flow on an elastic fiber~\cite{duprat2015fluid}, with particular focus on the transport~\cite{liu2018morphological} and deformation~\cite{duprat2015microfluidic} of slender elastic fibers by the flow~\cite{DuRoure2019dynamics}. 

Two-way flow-structure coupling has been studied in the case of deformable tubes, in which elastic deformations of the tube walls led to strong modifications of the fluid flow. Wall elasticity effects were shown to stabilize the flow distribution into a bifurcation~\cite{baroud2006propagation} and to suppress the emergence of viscous fingering in a Hele-Shaw cell~\cite{pihler2013modelling}. More extreme cases emerged in the case of air-liquid flows within flexible tubes, where surface tension led to a complete airway closure~\cite{hazel2005surface}. These effects of wall elasticity of the fluid flow have been used to explain sap flow in green plants~\cite{jensen_sap_2016} or to create soft valves for technological applications~\cite{kim_bi-polymer_2007,louf_bending_2020,chappel_review_2020}.

A different class of problems for which the two-way coupling can lead to extreme modifications of the flow consists of the case of a soft particle being pushed into an orifice. This problem is encountered in many microfluidic applications, such as during the flow and encapsulation of hydrogels~\cite{abate_beating_2009,klein_droplet_2015} or for the characterization of cells and other soft materials~\cite{luo_constriction_2014, elias_microfluidic_2020, preira_single_2013,tlili_microfluidic_2022,zhang_particle_2018,xu_micromechanics_2022}. It is also closely related to clogging or sieving particles~\cite{harth2020intermittent,dressaire_clogging_2017,agbangla2014collective,tao2021,hong2022,alborzi2022soft,yoon_clogging-free_2016}. In many of these cases a spherical particle is pushed into a rectangular slit, which allows for a leakage flow around the particle~\cite{luo_constriction_2014,preira_single_2013,khan_passage_2017,zhang_particle_2018}. But in spite of the interest for applications, the physics that determines the equilibrium between leaky flow and particle deformation has not been explored. Here we describe this nonlinear relationship and explain the low and high deformation regimes that are encountered.

\begin{figure}[!ht]
    \centering
    \includegraphics[width=0.8\columnwidth]{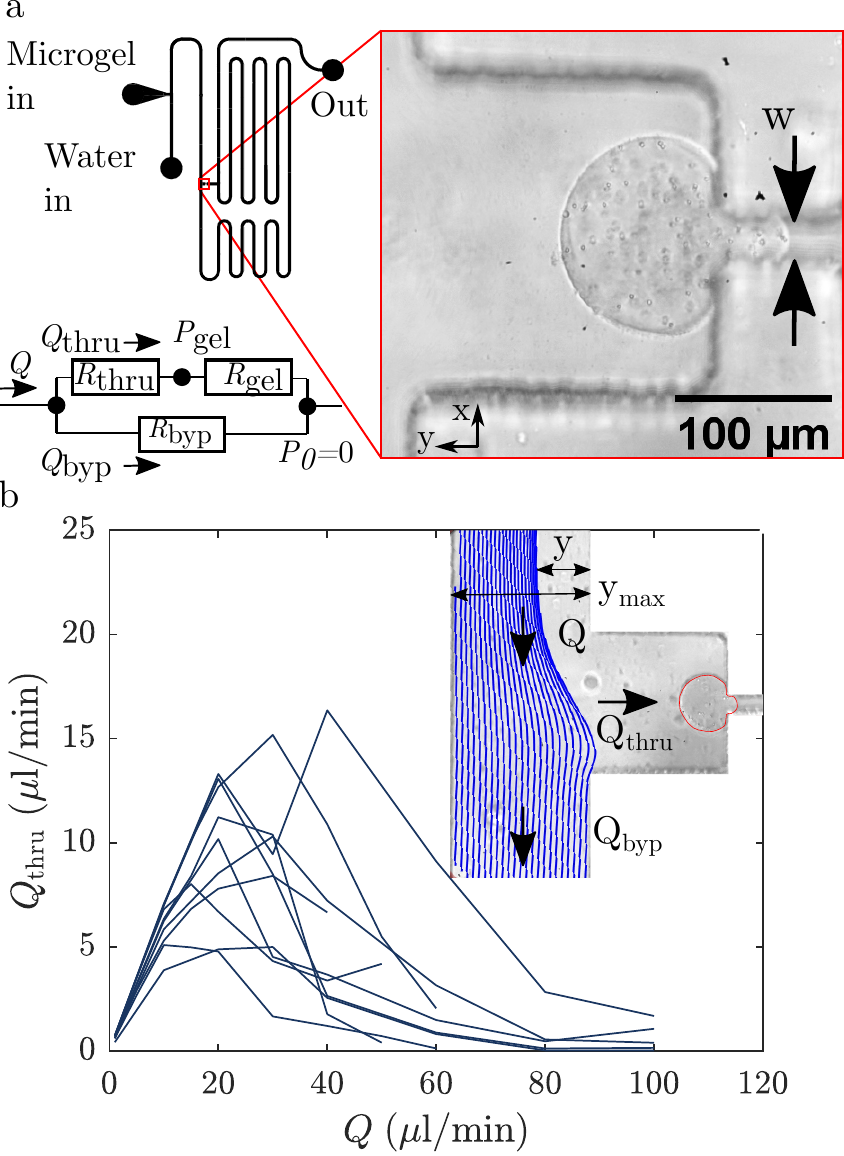}
    \caption{(a) Sketch of the microfluidic device. The flow rate is imposed at the inlet and divides into a thrupass and bypass channels. The micrograph shows a gel bead trapped in a slit of width $w=25$~\textmu m. A resistance diagram of the microfluidic channel is included. (b) The flow rate $\Qthru$ past the microgel increases to a maximum and then decreases to zero as the inlet flow rate $Q$ is increased. Each line represents a different trapped microgel. Young's modulus $E = 2.1\pm0.9\; \rm kPa$. Gel diameter: $d=83\pm3\; \mu \rm m$. Slit width: $w=25 \pm 5\; \mu \rm  m$. Channel height: $h=85\;  \mu \rm m$. Inset: reconstructed streamlines showing the flow division between the thrupass and bypass channels. The value of $y/y_{\rm max}$ provides a measure of the flow rates $\Qthru$ and $\Qbyp$.
    }
    \label{fig:MethodsLayout}
\end{figure}

The experimental setup consisted in a microfluidic device with two parallel channels: a thrupass line with a narrow slit to trap microgel beads, and a bypass line of width $y_{\rm max}$, which provided a known hydraulic resistance, see Fig.~\ref{fig:MethodsLayout}. The flow rate $Q$ at the inlet was imposed with a syringe pump (Nemesys, Cetoni GMBH, Germany), and streamlines of the flow in the device were reconstructed by tracking fluorescent particles (0.5 $\mu \rm m$ Polybead, Polysciences Inc., Washington Pa) with PIVLab~\cite{thielicke_pivlab-_2014,guermonprez_flow_2015}, see inset in Fig.~\ref{fig:MethodsLayout}b. The flow rate $\Qthru$ in the thrupass channel  was obtained by measuring the position $y$ of the separatrix streamline, shown in the inset of Fig.~\ref{fig:MethodsLayout}b. The separatrix divides the fluid going to the thrupass and the bypass channels, according to:
$\Qthru={\int_0^y U(y)dy}/{\int_0^{y_{\rm max}} U(y) dy}Q$, where $U(y)$ is the theoretical flow speed in a rectangular channel of a height $h$ and width $y_{\rm max}$~\cite{vanapalli_scaling_2007}. 

In the absence of a gel bead, increasing the flow rate $Q$ at the inlet led to an increase of the flow rate $\Qthru$, with the ratio between $\Qthru$ and $Q$ being given by the ratio of the hydrodynamic resistances between the thrupass and bypass channels. When a gel bead was trapped by the narrow slit, increasing the flow rate $Q$  at the inlet led to a non-monotonic behavior of the flow rate $\Qthru$ through the thrupass channel. At first, increasing $Q$ led to an increase in $\Qthru$, until a maximum value of $\Qthru$ was reached; increasing $Q$ beyond a critical value then led to a decrease in the flow rate $\Qthru$, see Fig.~\ref{fig:MethodsLayout}b. At high enough values of $Q$, the gel bead plugged the thrupass channel entirely and all the flow was redirected to the bypass channel. This non-monotonic relationship between $\Qthru$ and $Q$ reveals a strong non-linear hydrodynamic resistance added by the gel bead.

\begin{figure}[!htbp]
    \centering
    \includegraphics[width=\columnwidth]{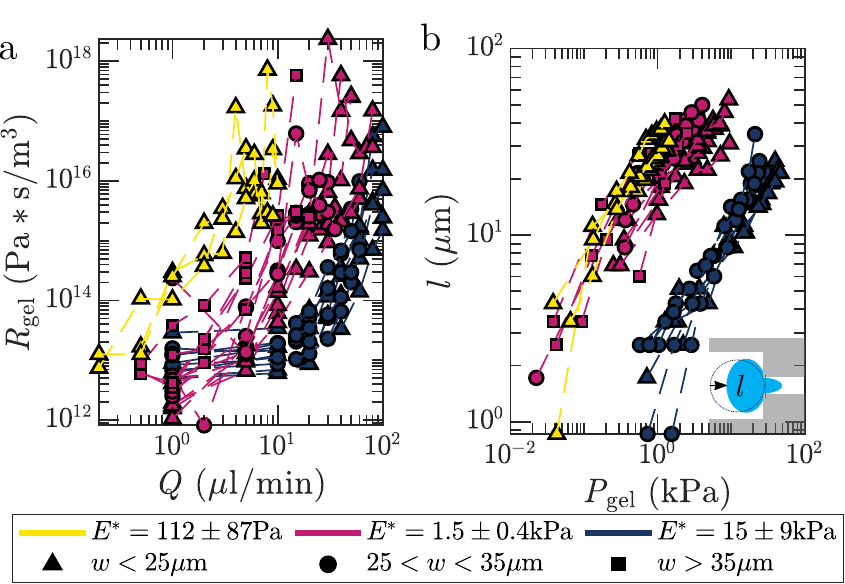}
    \caption{
    (a) Hydrodynamic resistance due to the gel $\Rgel$, as a function of the imposed flow rate $Q$. Note that the resistance increases faster for softer gels than stiffer gels.
    (b) Microgel displacement as a function of the applied pressure. Each curve represents data collected on a single microgel.
    }
    \label{fig:rawdata}
\end{figure}

Experiments were repeated for poly(ethylene glycol) (PEG) beads of diameters $d = 80 \; - 145 \;\mu \rm m$. The beads were made using droplet flow lithography by adding PEG-diacrylate (PEG-DA) to a PEG solution and photo-polymerizing it in a first microfluidic device~\cite{dendukuri_controlled_2005}. The beads were stored off-chip until they were re-injected into the current device. The microgel stiffness was controlled by modulating the ratio of PEG to PEG-DA. The values of the equivalent Young's moduli were obtained using a microindentation technique~\cite{guillou_dynamic_2016} and spanned two orders of magnitude: $E^{\star} = 100 \; \rm Pa$ to $ 21 \; \rm kPa$, where $E^{\star}=\frac{E}{{1-\nu^2}}$, $E$ is the conventional Young's modulus and $\nu$ is the Poisson ratio. This modulus is measured directly by microindentation and accounts for contact loading rather than simple compression/extension. Three different microfluidic traps were used, ranging within  $w = 15 - 45 \; \mu \rm m$, and microchannel heights $h$ chosen to approximately match microgel diameters: $h=d \pm 10 \%$.

At low Reynolds number, the pressure $\Delta P$ across a channel is proportional to the flow $Q$ in the channel and the hydrodynamic resistance $R$ of the channel: $\Delta P= Q R$~\cite{bruus2007theoretical}. Calling $\Rthru$ and $\Rbyp$ the known hydrodynamic resistances of the thrupass and bypass channels in the absence of gel and $\Rgel$ the added hydrodynamic resistance of the gel, we therefore have: $\frac{\Qbyp}{\Qthru} =\frac{(\Rthru + \Rgel)}{\Rbyp}$, which provides a way to compute the resistance $\Rgel$ added by the gel, as well as the pressure drop across the gel $P_{\rm gel} = \Rgel\Qthru$ (see Fig.~\ref{fig:MethodsLayout}a). 

The value of $\Rgel$ increased dramatically with flow rate, spanning nearly 6 orders of magnitude when the flow rates covered 2 orders of magnitude (see Fig.~\ref{fig:rawdata}a). The rate of this increase depended on the gel elasticity: stiffer gels ($E^{\star} \approx 10^4\; \rm Pa$) led to a slower increase in resistance than softer gels ($E^{\star} \approx 10^2\; \rm Pa$). The increase in resistance was associated with a displacement of the microgel as it deformed and penetrated into the slit. This displacement was quantified by measuring the distance $l$ traveled by the back of the gel, i.e. upstream of the slit, with respect to its position in the absence of flow (see inset in Fig.~\ref{fig:rawdata}b). The value of $l$ increased with the pressure drop across the gel $P_{\rm gel}$, see Fig.~\ref{fig:rawdata}b. For a given pressure $P_{\rm gel}$ across the gel bead, the softer gels penetrated more into the slit, and even more so when the gap width was larger, see Fig.~\ref{fig:rawdata}b. 

\begin{figure}[!htbp]
    \centering
    \includegraphics[width=\columnwidth]{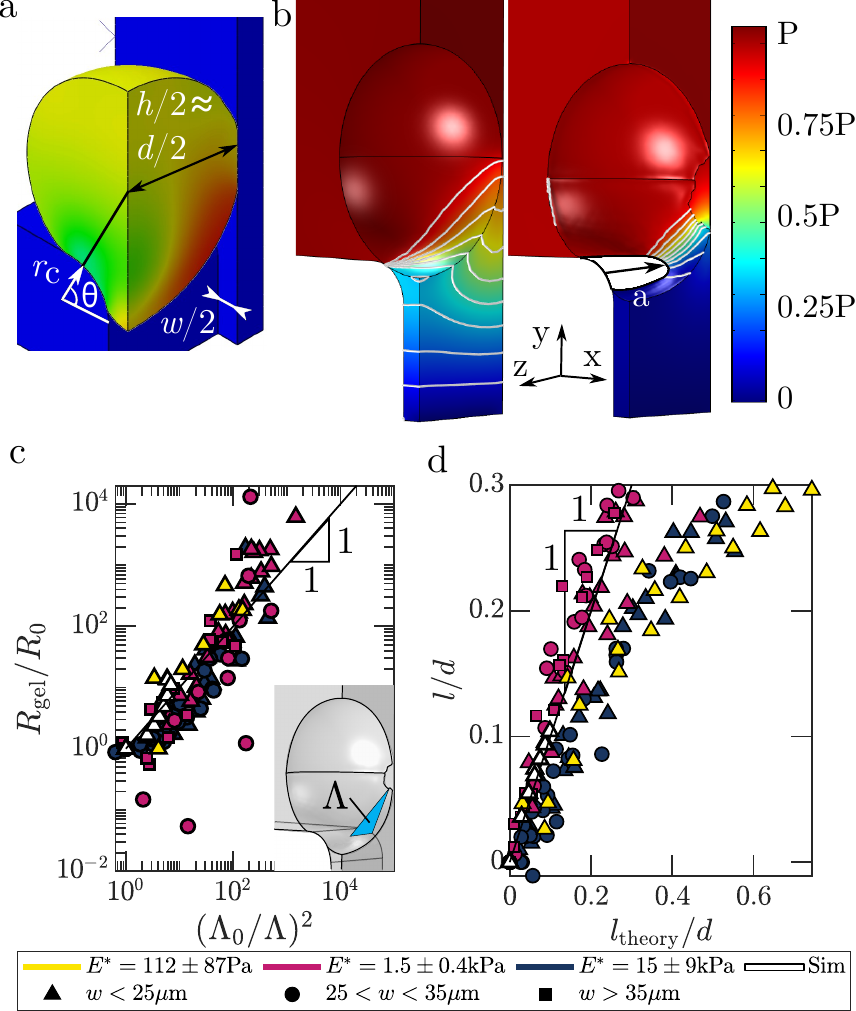}
    \caption{
    (a) A quarter section of the deformed microgel and trap for an applied pressure $P=4\; \rm kPa$. Color represents relative nodal displacement in the microgel.
    (b) Simulation of the flow around an undeformed gel (left), and a gel deformed under a simulated pressure of $P_{\rm gel}=4\; \rm kPa$ (right). Color indicates the pressure normalized by the inlet pressure $P$. Isobars are shown in white. Note how the deformed microgel focuses the pressure drop into a narrow area. $x$ and $y$ directions represent the imaging plane.
    (c) The measured gel hydrodynamic resistance $\Rgel$ as a function of the area $\Lambda$ of the triangle through which the flow must pass (see inset). $\Rgel$ and $\Lambda$ are respectively normalized by the resistance to flow $R_0$ and the area $\Lambda_0$ in the absence of gel deformation.
    (d) A comparison of the measured normalized deformation of the microgel $l/d$ with its predicted deformation based on solid contact, $l_{\rm theory}/d$. The graphs include experimental (color) and simulation results (white).
}    \label{fig:Contactfit}
\end{figure}

To understand how the flow forces the hydrogel to deform, plug the slit and increase the hydrodynamic resistance, simulations of a soft particle deforming into a slit were performed on a quarter-setup using the software Abaqus. The solid deformations were modeled by simulating an initially spherical gel bead that was subjected to a negative pressure inside the slit. The resulting deformed gel geometry (Fig.~\ref{fig:Contactfit}a) was exported to the software COMSOL Multiphysics and the flow field in the channel containing the deformed gel was simulated, providing the pressure distribution in the fluid everywhere in the device. The uncoupled solid and fluid simulations were then iterated once by updating the gel shape and then the pressure field. 

In the absence of deformation, the pressure drop in the channel occurred both around the spherical gel particle and in the slit, as shown by the isobars in the fluid simulations of Fig.~\ref{fig:Contactfit}b (left panel). When the gel deformed however, the pressure drop was focused over a short distance upstream of the slit, corresponding to the region where the flow was focused through a narrow gap, whose cross-sectional area we denote $\Lambda_{\rm exp}$ (Fig.~\ref{fig:Contactfit}b, right panel). This concentration of the pressure drop indicates that the resistance to flow was due not to the finger of gel elongating into the slit but rather to the bulb-like part of the gel that obstructed the fluid flow upstream of it. 

The physical reason behind the focusing of the pressure drop can be understood by measuring the dependence of $R_{\rm gel}$ on $\Lambda_{\rm exp}$. This was done by first measuring the area $\Lambda_{\rm sim}$ of this spacing on all numerical simulations (see inset in Fig.~\ref{fig:Contactfit}c). A geometric argument (see Supp. Mat. for details) yields a relation between $\Lambda_{\rm sim}$ and other geometrical parameters:
\begin{equation}
    \Lambda_{\rm sim} = \left(\frac{d}{4}-\frac{a}{2}\right)\sqrt{\left(\frac{d}{2}-l-\frac{a}{2}\right)^2+\left(\frac{w}{2}+\frac{r_c}{\sqrt{2}}\right)^2},
    \label{eq:AreaTriangle}
\end{equation}
where $d$ is the bead diameter, $w$ is the width of the slit, $r_c$ is the radius of curvature of the corner of the slit (see Fig.~\ref{fig:Contactfit}a), and $a$ is in effect the farthest extent of the contact area between the gel and the corner of the slit (see Fig.~\ref{fig:Contactfit}b, right panel).

Because $a$ is in a direction perpendicular to the imaging plane, it is inferred by considering the contact mechanics problem of a spherical elastic bead, of diameter $d$ and elastic modulus $E^{\star}$, being pressed with a pressure $P_{\rm gel}$ on top of two cylinders of equal radius $r_c$ and spaced by $w$~\cite{barber_contact_2018}. The displacement of the gel $l_{\rm theory}$ due to this forcing was found to relate to the major axis $a$ of the ellipsoidal contact between the microgel and the trap, modulated by a combination of the geometrical parameters of the problem (see Supp. Mat. for full derivation):
\begin{dmath}
    l_{\rm theory} = \frac{2a^2}{d\sin{\theta}}\mathcal{F}_{1}(r_c,d),
    \label{eq:penetrationLength}
\end{dmath}
where $\mathcal{F}_{1}(r_c,d)$ is a function that describes the shape of contact (see Supp. Mat.) and $\theta$ is the angle created between the microgel, the corner and the slit (see Fig.~\ref{fig:Contactfit}a).

The experimental area $\Lambda_{\rm exp}$ of the spacing between gel bead and sidewalls was similarly estimated from Eq.~\eqref{eq:AreaTriangle}, by measuring the geometric parameters $d,w,r_c$ and $l$. Using Eq.~\eqref{eq:penetrationLength}, we were able to estimate $a$ at each imposed flow rate as well. 

Given the above geometric measurements it is now possible to determine the scaling of $\Rgel$ with the interstitial space $\Lambda$. Calling  $R_0$ and $\Lambda_0$  the values of the parameters in the absence of gel deformation, \emph{i.e.} at the lowest flow rate tested, we plot $\Rgel/R_{0}$ as a function of $(\Lambda_0/\Lambda_{\rm exp})^2$ in  Fig.~\ref{fig:Contactfit}c. 

The plot shows an excellent collapse for all experimental and numerical results and indicates that $\Rgel$ scales as $\Lambda_{\rm exp}^{-2}$. The collapse of the measurements of $\Rgel$ on a single master curve shows that the resistance to flow is indeed due to the deformation of the soft solid upstream of the slit which in turn determines the size of the gap that the flow must go through. Moreover, the rapid increase of $\Rgel$ makes it the dominant source of pressure drop compared with a the other sections in the rest of the microchannel. 

In turn the added resistance due to flow focusing couples back to modify the shape of the microgel. This determines the values of $l$ and $a$, which are related together by Eq.~\eqref{eq:penetrationLength}. Contact mechanics modeling shows that $a$ depends on the ratio $P_{gel}/E^\star$ and a combination of the geometric parameters, which yields an implicit relationship between mechanical and geometric effects:
\begin{align}
    \left(\frac{a}{d}\right)^3 = & \frac{3P_{\rm gel}}{E^\star}\; \times \notag \\
    & \left[
    \frac{aw+\frac{2}{3}(\frac{d}{2}-a)[w+r_c(1-\cos{\theta})]}{4\pi d^2\sin{\theta}}\right] \mathcal{F}_{2}(r_c,d)
    \label{eq:MajorRadius}
\end{align}
where $\mathcal{F}_{2}(r_c,d)$ is another function of the shape of contact with the corner (see Supp. Mat.).

The expected theoretical value of the rescaled gel elongation $l_{\rm theory}/d$ was calculated from Eqs.~\eqref{eq:penetrationLength} and~\eqref{eq:MajorRadius} and compared to the experimentally and numerically computed values of elongation, as shown in Fig.~\ref{fig:Contactfit}d. Again the theory collapses the data onto a master curve, particularly for small deformations for which contact mechanics are expected to apply. 

It is now possible to quantitatively understand the relationship between the pressure $P_{\rm gel}$ on the trapped soft bead and the flow $\Qthru$ it lets pass through. Combining Eqs.~\eqref{eq:AreaTriangle}-\eqref{eq:MajorRadius}, we find the following relationship between the normalized flow rate through the slit $\mathcal{Q}={Q_{\rm thru}R_0}/{E^{\star}}$ and the normalized pressure drop $\mathcal{P} = P_{\rm gel}/E^{\star}$ across the gel:
\begin{multline}
\label{eq:QthruModel}
    \mathcal{Q} =
    C_1 \mathcal{P} \left(1-C_2\mathcal{P}^{\frac{1}{3}} \right)^2 \left(\left(1-C_3\mathcal{P}^{\frac{2}{3}}\right)^2+C_4^2\right)
\end{multline}
where the $C_i$ are geometric parameters dependent upon $w/d$ and $r_c/d$ (see Supp. Mat.).

\begin{figure}[!h]
    \centering
    \includegraphics[width=0.8\columnwidth]{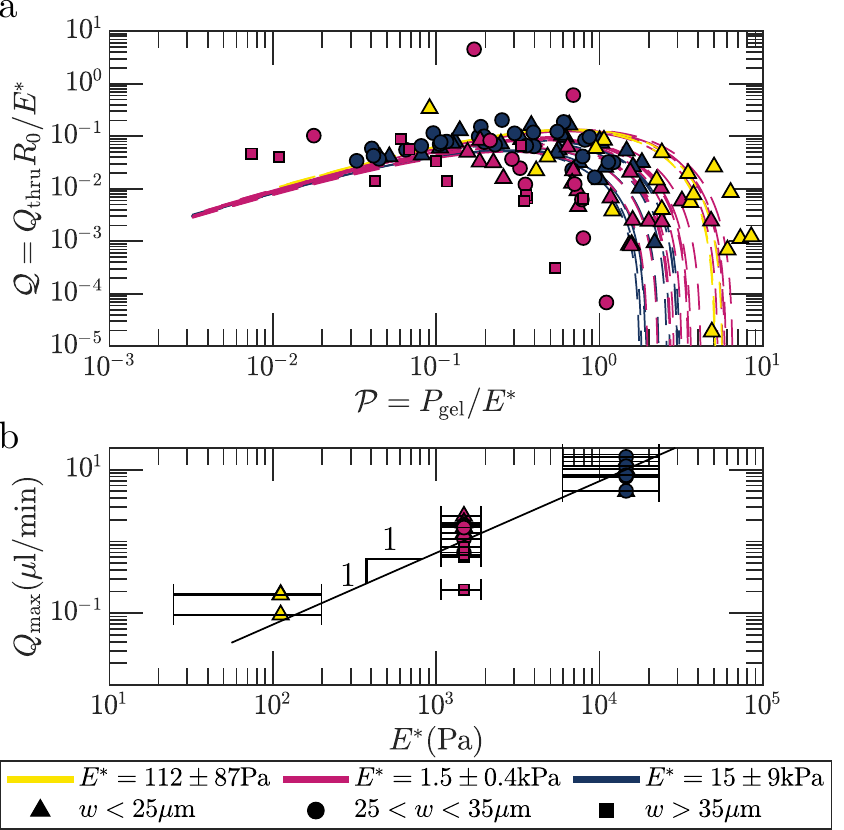}
    \caption{
    (a) Comparison of flow rate through the obstructed thrupass channel, shown in points, with the predicted flow rate, shown as dashed lines. Normalized flow rates, $\mathcal{Q}$, are given as a function of the normalized pressure $\mathcal{P}$ for the full range of pressures tested experimentally. (b) Maximum flow rate $Q_{\rm max}$ past the microgel as a function of the Young's modulus $E^{\star}$ of the trapped bead. Points: Experimental data. Error bars represent 1 standard deviation. Line shown represents the linear best fit.
}
    \label{fig:valveExplained}
\end{figure}

From Eq.~\eqref{eq:QthruModel} we can predict the evolution of $\Qthru$ as the driving pressure is increased. Experiment and theory are in good agreement, see Fig.~\ref{fig:valveExplained}a. For small values of $\mathcal{P}$ the pressure does not lead to significant deformation of the microgel, such that further increasing $\mathcal{P}$ leads to a nearly linear increase in $\mathcal{Q}$. When the pressure is increased beyond $\mathcal{P}\approx 1$ ($P_{\rm gel}\approx E^{\star}$) the gel bead deforms, which increases the value of $R_{\rm gel}$ and leads to a decrease in the flow rate through the slit. The transition between the two regimes depends on the slit geometry and shifts to lower values of $\mathcal{P}$ for larger values of $w$, indicating that wider slits are easier to plug than thinner slits. 

Finally, Eq.~\eqref{eq:QthruModel} shows that the dimensionless flow rate $\mathcal{Q}$ depends only on the dimensionless pressure $\mathcal{P}$ and on the geometry of the slit, through the parameters $C_1-C_4$. For a given geometry it follows that $\mathcal{Q}$ traces a unique curve, of  parameter $\mathcal{P}$, whose maximum value $\mathcal{Q}_{\rm max}$ thus only depends on the geometry of the slit. Therefore the maximum dimensional flow rate ${Q}_{\rm max} = \mathcal{Q}_{\rm max} E^{\star}/R_0$ scales linearly with the Young modulus of the gel $E^{\star}$: a soft gel deforms right away and plugs the channel at low pressures, while a stiff gel allows higher flow to go through. This linear increase is confirmed by comparing the largest measured flow rate with the prediction of $Q_{\rm max}$, as shown in Fig.~\ref{fig:valveExplained}b.

The results presented here can impact several applications, beyond their scientific interest. They provide a physical basis to understand the encapsulation of soft beads in droplets, which has emerged as an important microfluidic technology~\cite{zilionis_single-cell_2016}. In a different operation regime, the strong nonlinear relationship between pressure and flow rate can lead to the design of microfluidic nonlinear flow elements, such as check-valves or flow limiters. These devices play an important role in ensuring the robustness of fluidic circuits and protecting against surges. The analysis above shows that the maximum allowable flow rate scales linearly with $E^\star$, thus providing a simple design rule. Finally, the design of the slit presented here can also serve to measure the elastic modulus of soft materials, similarly to a micropipette aspiration device. 

\section*{Acknowledgements:} The authors acknowledge microfabrication assistance of Caroline Frot at Ecole Polytechnique and the Biomaterials and Microfluidics platform platform at Institut Pasteur. Useful discussions with Hiba Belkadi and Mahdi Daei Daei are also acknowledged. 

\bibliographystyle{apsrev4-2}
\bibliography{MyLibrary2372021.bib,alt_references.bib}

\end{document}